\def\comment#1{}
\newcommand{\bo}[1]{~\!\framebox[.71em]{\rule[-.0em]{0em}{.09em}{$\scriptsize{#1}$}}}
\def\Box{\raisebox{1.04mm}{$\bo{}$\hspace{1pt}}}
\documentstyle[eqsecnum,aps,epsf]{revtex}
\def\beqa{\begin{eqnarray}}
\def\eeqa{\end{eqnarray}}
\def\be{\begin{equation}}
\def\ee{\end{equation}}
\begin{document}
\title{Cosmology  with Curvature-Saturated Gravitational  Lagrangian
$R/\sqrt{1 + l^4 R^2}$ }
\author{Hagen Kleinert\thanks{kleinert@physik.fu-berlin.de,
http://www.physik.fu-berlin.de/\~{}kleinert}
and Hans-J\"urgen Schmidt\thanks{hjschmi@rz.uni-potsdam.de,
http://www.physik.fu-berlin.de/\~{}hjschmi}}
\address{Institut f\"ur Mathematik, Universit\"at Potsdam,
PF 601553, D-14415 Potsdam, Germany,\\
Institut f\"ur Theoretische Physik, Freie Universit\"at Berlin,
Arnimallee 14, D-14195 Berlin, Germany}
\date{\today}
\maketitle
\begin{abstract}
We  argue that the Lagrangian for gravity should
 remain bounded
 at large curvature, and
 interpolate between  the
weak-field tested Einstein-Hilbert Lagrangian
 ${\cal L}_{\rm EH} =  R /16 \pi G$
and a pure cosmological constant for  large $R$  with the
{\bf  \underline{c}}urvature-{\bf  \underline{s}}aturated ansatz
$ {\cal  L}_{\rm cs}={\cal L}_{\rm EH}/ \sqrt{1+l^4 R^2}$, where
$l$ is a length parameter expected to be  a few orders of magnitude
above the Planck length.
The curvature-dependent effective gravitational constant
 defined by $d{\cal L}/dR  = 1/16 \pi G_{\rm eff}$  is
 $G_{\rm eff} = G  \sqrt{1+l^4 R^2}^{~ 3} $,
 and tends to infinity for large $R$, in contrast to
 most other approaches where $G_{\rm eff} \to 0$.
The theory possesses neither ghosts nor tachyons, but it fails to be
 linearization 
stable. 
 In
a curvature saturated cosmology,
the coordinates with
$
ds^2 = a^2 \left[{da^2}/{B(a)} - dx^2 - dy^2- dz^2
\right]
$
are most convenient
since the curvature scalar becomes a linear
 function of $B(a)$. Cosmological solutions with 
a  singularity of type $ R \to \pm  \infty $  are possible 
 which have a bounded energy-momentum tensor everywhere; 
 such a behaviour is excluded  in Einstein's theory.
 In synchronized time, the metric is given by
\be
 ds^2 = dt^2 - t^{6/5}(dx^2 + dy^2+ dz^2).
\ee
On the technical side we show that
 two different  conformal transformations make ${\cal  L}_{\rm cs}$
 asymptotically  equivalent to the Gurovich-ansatz ${\cal  L}= \vert R \vert
^{4/3}$ on
the one hand, and to Einstein's theory with a minimally coupled
 scalar field with self-interaction on the other.
 %but  fails to be linearization stable.
\end{abstract}

\section{Introduction}
\setcounter{equation}{0}

According to an old idea by Sakharov \cite{1},
the gravitational properties of
spacetime are
 caused by the bending stiffness
of all quantum fields in a spacetime of scalar curvature $R$.
This idea of induced gravity has inspired many subsequent
theories of gravitation, from
Adler's \cite{3} proposal to consider Einstein gravity as a symmetry
 breaking effect in quantum field theory
 to the modern induced gravity derived from string fluctuations
\cite{3a}.
Whatever the precise mechanism, any induced gravity
will lead to a Lagrangian which is bounded at large
$R$, and may also go to zero. The latter case would be analogous to the elastic
stiffness
of solids, which is constant for small distortions,
but vanishes after the solid  cracks.

In this paper  we investigate the physical consequences of a simple
Lagrangian which goes to a constant at large $R$, thus
interpolating between the Einstein-Hilbert Lagrangian for small
$R$ and a pure cosmological constant for large $R$.
This Lagrangian will be referred to as {\em curvature-saturated\/}
and reads
\begin{equation}
{\cal L}_{\rm cs} = \frac{1}{16 \pi G} \frac{R}{ \sqrt{1 + l^4 R^2} }.
\label{@CS}\end{equation}
The length parameter $l$ may  range from an order of the Planck
length $l_{\rm P}$ or a few orders of
magnitude larger than $l_{\rm P}$.  Applying standard methods and those of
Refs.~\cite{4,5,6,7,8}, we shall derive the cosmological
consequences of the saturation and
compare our ansatz with others.

\bigskip
 
One of the motivations for a renewed interest in a more detailed %%New
 consideration of cosmology with non-linear curvature terms
 comes from M-theory, see   
  Ref. \cite{8a} ``Brane new world''. In \cite{8a} a  conformal anomaly  
 is considered, which turns out to have analogous consequences 
 as  Starobinsky's anomaly-driven inflation with $R$--  and $R^2$-terms, 
 see e.g. Refs. \cite{8b} for the older results.  Ref. \cite{8c}  contains 
  the latest results concerning the effective $\Lambda$-term in such models.

\bigskip 
 
Our own direct motivation to tackle the model discussed below was as follows: %%New
 We tried to make the analogy proposed in \cite{1} more closer than done by 
others; the analogy with solid state physics is this one: For small forces, the
 resistance to   bending is proportional to this force, but after a certain threshold --
 defined by cracking   the solid -- the resistance vanishes.

A similar line of reasoning was deduced in Ref. \cite{8d}: There the %%New
 finite-size effects
  from the closed Friedmann universe to the quantum states of fields have
 been calculated. Instead of continuous distribution of the energy levels of
 the
  quantum fields, one has a discrete spectrum.
 Qualitatively, the result is: If the radius $a$ of the spatial part
 of spacetime shrinks close to zero, which is almost the same as very large
$R$,
 then the spacings between the energy levels become larger and larger, and
after
 a certain threshold, all fields will be in the ground state.
 This behaviour shall be represented by an effective action. The concrete 
form of the corresponding effective Lagrangian is not yet fully determined
(that shall be the topic of later work), but preliminarily we found out that 
 the behaviour for large $R$ will quite probably  be of a Lagrangian bounded by 
 a special effective $\Lambda$; so we have chosen one of the easiest analytic functions 
possessing this large-$R$ behaviour  together with the correct weak-field shape. 

\bigskip

The paper is organized as follows: In Sec. II we calculate
 the consequences of the effective Lagrangian
${\cal  L}_{\rm cs}$.

In Sec.~III we investigate the consequences of the $R$-dependence of
the effective gravitational constant defined by
\begin{equation}
\frac{1}{16 \pi G_{\rm eff}} \equiv
\frac{d {\cal L} }{dR} ,
 \label{@2}\end{equation}
which is
\be\label{1.3}
G_{\rm eff} = G  \sqrt{1+l^4 R^2}^{~3}
\ee
for ${\cal L} = {\cal L}_{\rm cs}$     and  tends to infinity as $R \to \pm
\infty$.

Then  we apply two different  conformal transformations to
${\cal L}_{\rm cs}$. One   of them, presented in Sec.~IV,
 makes ${\cal  L}_{\rm cs}$
 asymptotically  equivalent to the Gurovich-ansatz \cite{9}, \cite{10}
\be\label{1.1}
{\cal  L}= \frac{R}{16 \pi G} + c_1 \vert R \vert ^{4/3} .
\ee
 The other transformation, by the Bicknell theorem given in Sec.~V,
   establishes a conformal relation to Einstein's theory,
 with a minimally coupled
 scalar field. In the literature, see \cite{11} and the references cited there,
 only the second of these conformal
transformations has so far been used.
The physical consequences of these three theories are, of course,
quite different since
the metrics are not
related to each other by coordinate
transformations.

Our approach differs fundamentally from that
derived from the
{\em limiting curvature hypothesis\/} (LCH) %%New
 in Refs. \cite{12}, where the  gravitational Lagrangian reads
\be\label{1.2}
{\cal  L}=R + \frac{\Lambda}{2}
\left(  \sqrt{1 - R^2/\Lambda^2} - 1 \right)
\label{@bound}\ee
whose derivative with respect to $R$ diverges for $R \rightarrow  \Lambda $.
This divergence was supposed to prevent  a curvature singularity,
a purpose not completely reached
  by the      %%New
 model presented in the first of Refs. \cite{12}  
 because other curvature  invariants may still diverge. (Let us note for completeness:  
In the second of Refs. \cite{12}, a more detailed version of the LCH is %%New
presented  which covers also the bounding of the other curvature invariants; 
 it is restricted to isotropic cosmological models. For more general space-times
 one faces the problem that sometimes  a curvature singularity exists, but 
all polynomial curvature invariants remain bounded there.) 

In contrast to Eq. (\ref{@bound}), our
model {\em favors\/} high curvature values.

\bigskip

It turns out that the use of synchronized or conformal time is not
optimal for our problem. We therefore use a new time coordinate which we call
 {\em curvature time\/}
for the spatially flat Friedmann model. The general properties
of this coordinate choice are described in Sec.~VI.

In Sec.~VII  we study the
consequences of curvature-saturation  for
some cosmological models using the coordinates of Sec.~VI.
In Sec.~VIII, finally, we
summarize our  results and compare with the related papers \cite{13} to 
\cite{25}.

\section{Field Equations of Curvature-Saturated Gravity}%%sct. II
The curvature-saturated Lagrangian (\ref{@CS})
interpolates between the
Einstein-Hilbert
Lagrangian
\be %2.1
 {\cal L}_{\rm EH} = \frac{R}{16 \pi G},
\label{2.1}\ee
which is
experimentally confirmed
at weak fields,
and a pure cosmological constant
at strong fields
\begin{figure}[tb]
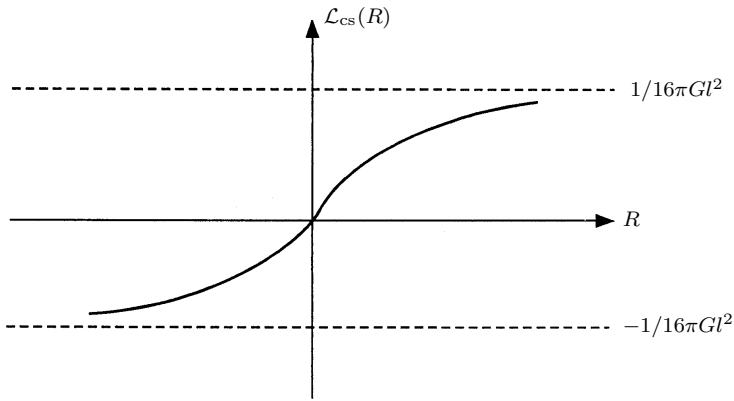

~\\
 \input s119fig4.tps
\caption[]{Curvature-Saturated
 Lagrangian %${\cal  L}_{\rm cs}$
  as a function of the curvature scalar $R$.
}
\label{@fig1}\end{figure}
\be\label{2.2}
{\cal L}  = \frac{\pm  1}{  16 \pi G l^2  }.
\ee
The $R$ dependence is plotted in Fig.~\ref{@fig1}.

The usual gravitational constant is obtained from the
derivative of the Einstein-Hilbert Lagrangian:
\be\label{2.3}
\frac{1}{16 \pi G} = \frac{d{\cal L}_{\rm EH}}
{dR}.
\ee
{}From our
 curvature-saturated
Lagrangian  (\ref{@CS})
we obtain, with this derivative,
the effective gravitational
constant   (\ref{1.3}).
The definition   (\ref{2.3})  is motivated as
follows: If one considers the Newtonian limit
for a general Lagrangian ${\cal L}(R)$
which may contain a nonvanishing
 cosmological constant,
the potential between two  point masses
contains a Newtonian $1/r$-part plus a Yukawa-like part $\exp (-r/r_Y)$
stemming from
the nonlinearities of the Lagrangian; the details are given in the
Appendix.
At distances much larger than $r_Y$, but much smaller than $1/\sqrt{R}$,
only the $1/r$-term survives, and the
coupling strength
of the $1/r$-term
is
given
by the effective gravitational  constant $G_{\rm eff}$.
For a recent version to
deduce such weak-field expressions, see Ref. \cite{20}.

For a general Lagrangian
${\cal L}(R)$ such as (\ref{@CS}),
the calculation of the field equation is somewhat
tedious, since the
Palatini formalism
which simplifies the calculation in Einstein's theory
is no longer applicable.
Recall that in this,  metric
   and the affine connection are varied independently,
the latter being identified with the
Christoffel symbol only at the end.

Here the following
indirect procedure leads rather efficiently to the correct field equations.
Let
\begin{equation}
{\cal L}'\equiv \frac{d{\cal L}}{dR},  \qquad
 {\cal L}'' \equiv \frac{d^2{\cal L}}{dR^2},
\label{2.3a}\end{equation}
and form
the covariant energy-momentum tensor of the gravitational field
which is given by the variational derivative of ${\cal L}$ with
respect to the metric $g_{ab}$:
\be\label{@5}
\Theta _{ab} \equiv \frac{2}{\sqrt {- g}} \frac{\delta {\cal L}
 \sqrt{- g}}{\delta g_{ab}},
\ee
 where $g$
denotes the determinant of $g_{ab}$.
For   dimensional reasons, $\Theta _{ab}$ has
the following structure
\be\label{@6}
\Theta _{ab} = \alpha {\cal L}' R_{ab} + \beta {\cal L}' R g_{ab} + \gamma
{\cal L} g_{ab}
+ \delta \Box {\cal L}' g_{ab} + \epsilon {\cal L}'_{;ab}
\ee
with the 5 real constants $\alpha \dots \epsilon$. These constants can be
uniquely
determined up to one overall constant factor  by
the covariant conservation law
\be\label{@7}
\Theta ^{ab}_{~~  ~ ;b} =0.
\ee
The overall factor is fixed by the Einstein limit $l\rightarrow 0$ of
the theory, where
 $\Theta _{ab} = ( R_{ab} - \frac{1}{2} R g_{ab})/8\pi G$.
In this way  we derive the following form of the covariantly conserved
energy-momentum tensor of the gravitational field
\be
\Theta _{ab} = \frac{1}{16\pi G}\left(2{\cal L}' R_{ab}- {\cal L} g_{ab}
+ 2 \Box {\cal L}' g_{ab}- 2 {\cal L}'_{;ab}\right).
\label{Theta}\ee
The calculation  is straightforward,
%\mn{hier hatten Sie noch die
%Zwischengleichungen angeben wollen.}
 if one is careful to distinguish
between
  $(\Box {\cal L}')_{;a}$ and $\Box ({\cal L}'_{;a})$, which differ by a
 multiple of the curvature scalar.

Inserting
our curvature-saturated Lagrangian
(\ref{@CS}) into
(\ref{2.3a})
and
omitting the subscript, we have
\be\label{4.2}
{\cal L} =  \frac{R}{2}  \left(1+l^4 R^2 \right)^{- 1/2} ~ ,  \qquad
{\cal L}' = \frac{d{\cal L}}{dR} = \frac{1}{2} \left(1+l^4 R^2 \right)^{- 3/2},
\ee
and find from
(\ref{Theta})
\be\label{4.3}
\Theta_{ab} =
\frac{1}{8\pi G}\left\{
 \frac{R_{ab}}{\left( 1 + l^4R^2 \right)^{3/2}} -
 \frac{R g_{ab}}{2 \left( 1 + l^4R^2 \right)^{1/2}} +
 g_{ab} \Box \left[ \frac{1}{\left( 1 + l^4R^2 \right)^{3/2}}\right] -
\left[ \frac{1}{\left( 1 + l^4R^2 \right)^{3/2}} \right]  _{;ab}
\right\} .
\ee
Setting $l=0$ reduces this to $1/16\pi G$ times the
 Einstein  tensor. The trace of (\ref{4.3}) is
\be\label{@b}
\Theta_{a}{}^a=
\frac{1}{8\pi G}\left\{
\frac{R+ 2l^4 R^3}{\left( 1 + l^4R^2 \right)^{3/2} }-
3  \Box \left[ \frac{1}{\left( 1 + l^4R^2 \right)^{3/2}}
\right]
\right\} .
\label{@tr}\ee
According to Einstein's equation, $\Theta_{ab}$ has to be equal
to the energy momentum tensor of the matter $T_{ab}$, i.e.,
 $T_{ab}=\Theta_{ab}$.
Equation (\ref{@tr}) implies that in the vacuum,
the only constant curvature scalar is
 $R=0$, such that this model does not possess a de Sitter solution.
Further, we can see from Eq. (\ref{4.3}), that a curvature
singularity does not
necessarily imply a divergence of energy-momentum, but may be
compensated  by the infinity of $G_{\rm eff}$.

\section{Effective gravitational constant and weak-field behavior}%%sct. III
\begin{figure}[htb]
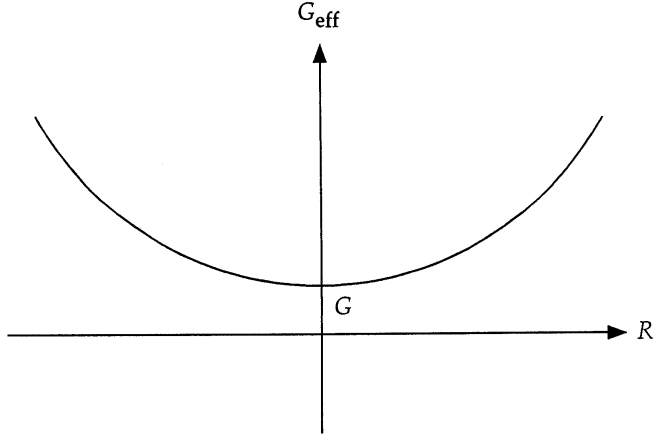

 \input s119fig2.tps
\caption[]{
Effective gravitational constant as a function of the curvature scalar.
}
\label{@fig2}\end{figure}

Let us compare
the effective gravitational constant $G_{\rm eff}$
of our curvature-saturated model
with those of
other models discussed
in the literature.
{}From (\ref{1.3})
we see that
 $G_{\rm eff}$
has the weak-field expansion
\be\label{3.1c}
G_{\rm eff} = G\left(1 + \frac{3}{2} l^4 R^2 + \dots\right) \,  ,
\ee
and the strong-field expansion
\be\label{3.2}
G_{\rm eff}
= G l^6 \vert R \vert ^3
\left( 1 + \frac{3}{2l^4 R^2} + \dots  \right).
\ee
The full $R$-behavior is plotted in Fig. \ref{@fig2}.

The weak-field expansion of ${\cal L}_{\rm cs}$ is given by
\be\label{@11}%\be\label{7.1}
{\cal L}_{\rm cs} = \frac{R}{16 \pi G \sqrt{1+l^4 R^2}}=  \frac{R}{16 \pi G} +
\sum_{k=1}^{\infty} ~b_k ~R^{2k+1}
\ee%\ee
 with real coefficients $b_k$, where
$ b_1 = -{l^4}/{32 \pi G}$.

As one can see, the quadratic term is absent, so that
the linearized field equation coincides with the linearized Einstein
 equation. Thus we encounter  neither ghosts nor tachyons; for details see 
Appendix B. %%New

There is, however, a price to pay for it.
The theory
has lost
linearization stability of the
solutions. This latter property has the following consequences: If one performs
 a weak-field expansion
\be\label{@12}
g_{ij} = \eta_{ij} + \sum_{m=1}^\infty ~\epsilon^m g^{(m)}_{ij}
\ee
around flat spacetime to solve the field equation, one
has to use the terms up to  the  order $m=2$  to get the complete
 weak-field part of the set of solutions. With  this peculiarity, we obtain
a well-posed Cauchy problem for the gravity theory following from the
Lagrangian ${\cal L}_{\rm cs}$.

%\subsection{ Comparison with  other Lagrangians}%III.2

Let us now compare our theory with others available in the literature. Let
%For ease of comparison with the models discussed earlier we give the
%analogous results for
\be\label{3.3}
{\cal L}_{\alpha,n}(R)= \frac{R}{16 \pi G}+ \alpha R^n
\ee
with some number $ n >1$ and constant $\alpha \ne 0$. In analogy with
Eq. (\ref{@2})
we calculate the effective gravitational constant from
\be\label{3.4}
 \frac{1}{16 \pi G_{\rm eff}} =
\frac{d{\cal L}_{\alpha,n}}{dR} = \frac{1}{16 \pi G} + \alpha n R^{n-1}
\ee
%which can be inverted to
such that
\be\label{3.5}
G_{\rm eff} = \frac{G}{1 + 16 \pi \alpha n
R^{n-1}G},
\ee
i.e., $G_{\rm eff} \to 0$ as $R\to \pm \infty$.
For $n=2$, more exactly: for all even natural numbers $n$,  we meet an
additional peculiarity that $G_{\rm eff}$ can diverge for finite values of $R$
already. Such values of $R = R_{\rm crit}$ are called {\em critical\/}
\cite{4}.
 For $n =2$ we get
\be\label{@d}
 R_{\rm crit} = - \frac{1}{32 \alpha \pi G},
\ee
and this is the region where $G_{\rm eff}$ changes its sign, as shown
in Figures 3 and 4.
At critical values of the curvature scalar, the Cauchy problem fails to be a
 well-posed one.

\begin{figure}[htb]
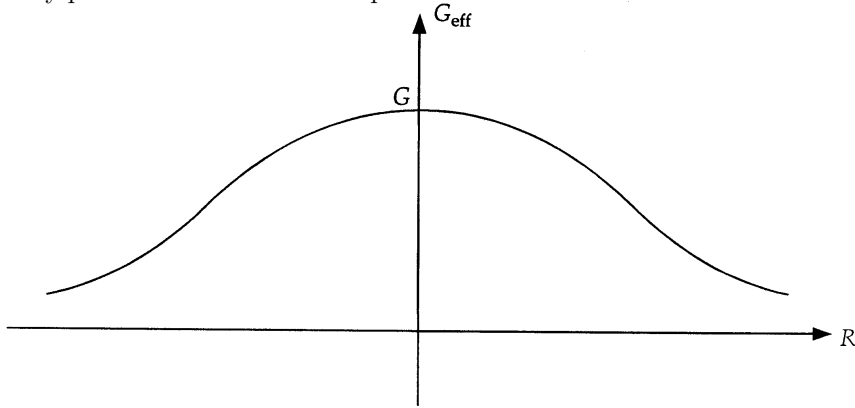

 \input s119fig3.tps
\caption[]{
Effective gravitational constant
 $G_{\rm eff}$ for ${\cal L}_{\alpha,3}$ with $\alpha > 0$
 as a function of $R$.
}
\label{@fig3}\end{figure}

\vspace*{1cm}

\begin{figure}[htb]
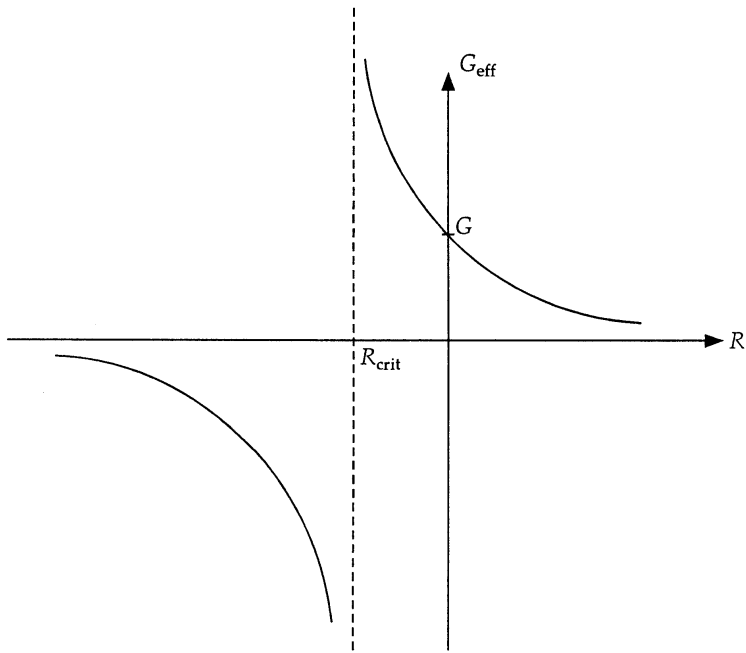

 \input s119fig1.tps
\caption[]{
Effective gravitational constant
 $G_{\rm eff}$ for ${\cal L}_{\alpha,2}$ with $\alpha > 0$
 as a function of $R$.
}
\label{@fig4}\end{figure}

\section{Conformal duality}%4.B

In Ref. \cite{8}, a duality transformation relating between different types of
nonlinear Lagrangians has been found. In the present notation it implies the
following
relation.
Let
\begin{equation}
\hat g_{ab}={\cal L}'^{2}g_{ab} %4.4
\label{4.1c}\end{equation}
be the conformally transformed metric
with ${\cal L}'\neq0$, which is fulfilled by our Lagrangian (\ref{@CS}).
Then the conformally transformed curvature scalar equals
\begin{equation}%4.5
\hat R = \frac {3R}{{\cal L}'^2}- \frac {4{\cal L}}{{\cal L}'^3},
\end{equation}
and the associated Lagrangian is %\mn{"dual" not defined}
\begin{equation}%4.6
\hat {\cal L} = \frac {2 R}{{\cal L}'^3} -  \frac {3 {\cal L}}{{\cal L}'^4}.
\label{4.6}\end{equation}
We easily verify that
$\hat {\cal L}' \, {\cal L}' = 1$.
Then one can prove that
$g_{ab}$ solves the vacuum field equation following from
${\cal L}(R)$ if and only if $ \hat g_{ab}$ of Eq. (\ref{4.1c}) solves
the corresponding equation for $\hat {\cal L}(\hat R)$
of Eq. (\ref{4.6}).

Example: For  ${\cal L}=R^{k+1}$ we find,
up to an inessential constant factor,
 $\hat {\cal L} = \hat R^{\hat k +1}$
 with $\hat k = 1/(2-1/k)$, such that for
a purely quadratic theory with ${\cal L}=R^2$,
 also $\hat {\cal L} = \hat R^2$.
For our curvature-saturated model ${\cal L} \to $ const.  we
 should expect a behavior with $k \to -1$, i.e., $\hat k \to 1/3$, this leads
to
$\hat {\cal L} \sim \hat R ^{4/3}$, which is  the Gurovich-model \cite{9}, cf.
Eq. (\ref{1.1}).

Let us study this  in more detail. To simplify the expressions we
 use, in this subsection only, reduced units with  $16\pi G=1$
 to best exhibit the fixed point $l=0$ of this transformation
making it an identity transformation
if applied
to Einstein's theory where $k=1$.
In the present
units,
 Eqs.  (\ref{4.2}) have to be multiplied by 2 and  become
\be\label{4.7}
{\cal L} = {R} \left(1+l^4 R^2 \right)^{- 1/2},~~~~~
\label{4.7b}
{\cal L}' = \frac{d{\cal L}}{dR} =  \left(1+l^4 R^2 \right)^{- 3/2} .
\ee
Inserting these into
(\ref{4.1c})--(\ref{4.6}),
  we obtain
\be\label{4.9}
\hat g_{ab} = \frac{g_{ab}}{(1+l^4R^2)^{\, 3}}
\ee
and
\be\label{4.10}
\hat R = - R(1+l^4R^2)^{\, 3} (1-4l^4R^2).
\ee
For small $R$ we have
\begin{equation}
\hat R = - R(1-l^4R^2 + \dots),
\label{@4.small}\end{equation}
and for large
$\vert R \vert $
\be\label{4.11}
\hat R = 4 l^{16} R^9 \left(1 + \frac{11}{4l^4 R^2} + \dots\right).
\ee
The inverse function $R(\hat R)$ of (\ref{4.10}) is not expressible in
closed form,
 but its small- and large-curvature expansion can be calculated from
 (\ref{@4.small}) and
 (\ref{4.11})
\be\label{4.12}
 R = - \hat R(1+l^4\hat R^2 + \dots),~~~~~~
R = \left(   \frac{\hat R }{4 l^{16}}   \right)^{\, 1/9 }
\left[1 - \frac{11}{36 l^4} \left( \frac{4l^{16}}{\hat R}
\right)^{\, 2/9} + \dots
 \right]
\ee
>From Eq. (\ref{4.6}) we see that
\be\label{4.13}
\hat {\cal L} = - R(1+l^4 R^2)^{\, 9/2} (1-3l^4 R^2)
\ee
where  $R(\hat R)$ has to be inserted.
For large $R$  we use the right-hand equation in (\ref{4.12}) and
obtain
the limiting behavior
\be\label{4.14}
\hat {\cal L} = 3l^{22} \left(\frac{\hat R}{4l^{16}} \right)^{\, 4/3} \,
\left[1 -
\frac{51}{6l^4} \left( \frac{ 4l^{16}}{\hat R}  \right)^{\, 2/9}
+ \dots
\right].
\ee

\section{Bicknell's theorem}%4.C

\begin{figure}[tb]
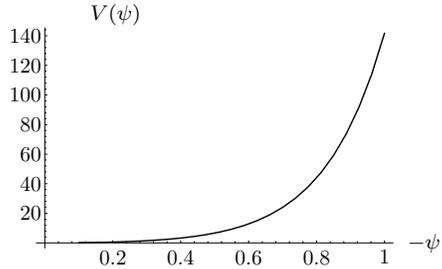

 \input psipot.tps
\caption[]{Potential $V(\psi)$ associated with curvature-saturated action via Bicknell's theorem.
}
\label{@figpsi}\end{figure}

Bicknell's theore \cite{21},
in the form described  in Ref. \cite{4},  relates
 Lagrangians
of the type
 (\ref{4.2})
to Einstein's theory coupled minimally
to a scalar field $\phi$ with a certain
interaction potential $\tilde V(\phi)$.
This Lagrangian is given by
\be\label{@50}
{\cal L}_{\rm EH}  + \frac{1}{2} \phi_{,i} \phi^{,i} - \tilde V(\phi) ~.
\ee
The relation of $ \tilde V(\phi)$ with ${\cal L}(R)$ is
expressed most simply by
defining a field with a different normalization
$\psi = \sqrt{2/3} \, \phi $, in terms of which
the  potential
 $\tilde V(\phi)= V(\psi) $
reads
\be\label{@51}
V(\psi)={\cal L}(R)e^{-2\psi} - \frac{R}{2} e^{-\psi},
\ee
with $R$ being the inverse function of
\be\label{4.16a}
\psi = \ln[2{\cal L}'(R)].
\ee
 The metric in the transformed Lagrangian
(\ref{@50}) is
\be\label{4.15}
\tilde g_{ab} = e^\psi g_{ab}.
\ee
For  our particular Lagrangian (\ref{4.2}) we have from (\ref{4.16a}):
\be\label{4.16}
\psi =  - \frac{3}{2} \ln(1+l^4R^2).
\ee
Now we restrict our attention to the range $R>0$ where $\psi <0$;
the other sign can be treated
analogously. Then (\ref{4.16}) is inverted to
\be\label{4.17}
R = \frac{1}{l^2} \sqrt{ e^{-2\psi/3} -1 }        ,
\ee
such that (\ref{@51})  becomes
\be\label{4.18}
V(\psi)=\frac{1}{2l^2}( e^{-5\psi/3} - e^{-\psi} )\sqrt{ e^{-2\psi/3} -1}.
\ee
In the range under consideration, this is a positive and
monotonously increasing function of $-\psi$ (see Fig.~\ref{@figpsi}),
with the large-$\Phi$ behavior
\be\label{@54}
V=\frac{1}{2l^2} e^{-2\psi} .
\ee
This is  the typical exponential potential for power-law inflation.
As mentioned at the end of Section II, no
exact de Sitter inflation exists. For $\psi \to 0$, also $V(\psi)\to 0$ like
$4\sqrt{2/3}\psi^{3/2}$.

\medskip

If $V(\psi)$ has a quadratic minimum at some $\psi_0$ with positive value $V_0=V(\psi_0)$,
then there exists a stable de Sitter inflationary phase.
As a pleasant feature, the potential  $V(\psi)$ has no maximum which
have given rise to tachyons.

\medskip

{}From Eq. (\ref{4.16}) one can see that for weak fields, $\psi \sim R^2$,
whereas
 a $R+R^2$-theory has  $\psi \sim R$. In other words: In our
model it is a better approximation
to assume
 the conformal factor $e^\psi$ to be approximately constant
for weak fields then in
 $R+R^2$-theories, since at the  level keeping  only terms linear in $R$
the two  metrics $g_{ab}$ and $\tilde g_{ab}$
 in (\ref{4.15}) coincide.

\medskip

\section{Friedmann models in curvature time}
The expanding spatially flat Friedmann model
may be parametrized  with the help of {\em curvature time\/} $ a > 0$ as follows:
\begin{eqnarray}
 ds^2 = a^2 \left[ \frac{da^2}{B(a)}  - dx^2 - dy^2 - dz^2 \right] ,
\label{1}\end{eqnarray}
where $B (a) $ is an arbitrary positive function determining
 $R$ as
\begin{eqnarray}
 R = -\frac{3}{a^3}  \frac{dB}{da},
\label{2}\end{eqnarray}
depending only on the first derivative of $B(a)$.
%which is the reason for the name ``curvature time'' for $t$.
This is a special feature of (\ref{1}) since, in general,
the curvature scalar depends on the
second derivative of the metric components.
Note also the
{\em linear \/}
dependence of $R$ on $B' \equiv {dB}/{da}$,
in contrast to
the usual nonlinear dependence of the curvature scalar on the first
derivative of the metric coefficients.

Let us recall some facts on Friedmann models
in curvature time and
exhibit the corresponding transformation
to synchronized time.
\subsection{From curvature time to synchronized time}
\label{@VA}%
The spatially flat Friedmann model in
synchronized time has the  metric
\begin{eqnarray}
 ds^2 = dt^2 - a^2(t) (dx^2 + dy^2 + dz^2).
\label{3}\end{eqnarray}
Metric (\ref{1}) goes over to metric (\ref{3}) via
\be
 dt = \frac{a \,d a}{ \sqrt{B(a)} }~~~,
\ee
such that
\begin{eqnarray}
  t = t (a) = \int \frac{ada}{ \sqrt{B(a)} }  .
\label{4}\end{eqnarray}
The inverse function $a (t)$ provides us with the desired
 transformation.
\subsection{From synchronized time  to curvature time}
Consider  $a(t)$ in an expanding model with
 \be
\dot a \equiv \frac{da}{dt} > 0.
\ee
Then we can invert $a (t)$ to $t(a)$, and have
\begin{eqnarray}
 B(a) = a^2 \left[ \dot a (t (a)\right] ^2   .
\label{5}\end{eqnarray}
>From this relation we understand
 why $R$ depends on the first
derivative of $B$ only: $B$ itself contains a
derivative of $a$, and $R$ is known to contain up
 to second order derivatives of $a(t)$.
\subsection{ Examples}

 Let $a(t) = t^n$, i.e. $t = a^{1/n}$, ~~$\dot a(t) = n t^{n-1}$,~~
 $\dot a (t(a)) = n a^{1-1/n}$. Then
  Eq.~(\ref{5})  yields
\begin{eqnarray}
 B(a) = n^2 a^{4-2/n}.
\label{6}\end{eqnarray}
 Let further $ a(t) = e^{Ht}$, $H = \mbox{const.} > 0$, $\dot a = Ha$. Then
\begin{eqnarray}
   B(a) = H^2 a^4 \quad .
\label{7}\end{eqnarray}
 Obviously, Eq.  (\ref{7}) is a limiting form
of Eq.~(\ref{6})
for
$ n \rightarrow \infty$. Equation~(\ref{1}) with $B(a)$ from (\ref{7}) represents
a vacuum solution of Einstein's theory with $ \Lambda $-term
where $ \Lambda = 3 H^2$, namely the de Sitter spacetime.

Let us also give some examples for the direct use the curvature time:
\begin{itemize}
\item[1.)] From Eq.~(\ref{2}) we see that
 $R = 0$ implies $B \equiv $ const., corresponding to
 $n = \frac{1}{2}$ in Eq. (\ref{6}), i.e., $ a = t^{1/2}$
in synchronized time.
%%%
This is the usual Friedmann radiation model.
\item[2.)] Also from Eq.~(\ref{2}), a constant $R$
$\neq 0$ implies
 $B = C_1 + C_2 a^4$ with
 constants $C_1$ and $C_2$,~ $C_2 \ne 0$.
\\
For $C_1 = 0, C_2 = H^2$, this represents the
de Sitter spacetime Eq.~(\ref{7}).
\item[3.)]
 The dust-model in synchronized coordinates is
given by $ a = t^{2/3}$, i.e.,
 with Eq.~(\ref{6}) we get
\begin{eqnarray}
B (a) = \frac{4}{9} a,
\label{@X1}\end{eqnarray}
such that $B' =$ const. Together with Eq.~(\ref{2}), this leads to
\begin{equation}
 R\, a^3 = \mbox{const,}
\label{@X2}\end{equation}
ensuring  mass conservation, because $R$ is proportional
to the mass density, and the pressure is negligible
for dust.
 \end{itemize}
\subsection{ The variational derivative}
For the metric (\ref{1})
we have
\begin{eqnarray}
  \sqrt{-g}  \equiv  \sqrt{- \det g_{ij}} =
 \frac{a^4}{ \sqrt{B} }.
\label{8}\end{eqnarray}
 The Lagrangian for Einstein's theory with $ \Lambda $-term reads
\begin{eqnarray}
{\cal L} = (R + 2  \Lambda ) \sqrt{-g} \quad .
\label{9}\end{eqnarray}
With (\ref{2}) and (\ref{5}) we get from (\ref{9})
\begin{equation}
{\cal L}  = \left(2  \Lambda - \frac{3B'}{a^3}\right)
 a^4 B^{-1/2}      .
\label{@X3}\end{equation}
The vanishing of the variational derivative
\begin{equation}
\frac{ \delta {\cal L}}{ \delta B} \equiv  \frac{\partial {\cal L}}{\partial B}
 - \left(\frac{\partial {\cal L}}{\partial B'}\right)' = 0
\label{@X4}\end{equation}
gives
$B = H^2 a^4 $ with $  \Lambda = 3 H^2$, i.e.,
the usual de Sitter spacetime.
No integration is necessary, since the derivative of $B$
cancels.
Intermediate expressions
are
\begin{equation}
\frac{\partial {\cal L}}{\partial B} = \left(2  \Lambda  -
  \frac{3 B'}{a^3}\right) \, a^4 \, \left(-\frac{1}{2}\right)
 B^{-3/2},
\label{@X5}\end{equation}
\begin{equation}
\frac{\partial {\cal L}}{\partial B'} = - 3 a B^{-1/2} ~  , \qquad
 \left(\frac{\partial {\cal L}}{\partial B'}\right)' =
 - 3  B^{-1/2} + \frac{3}{2}a B'  B^{-3/2}   .
\label{@X6}\end{equation}
\subsection{ Remaining coordinate-freedom}
Translations in $t$ do not change the form of the metric (\ref{3}).
 This freedom is related to the fact that the integration constant
in the integral (\ref{4}) remains undetermined; this coordinate freedom has no
analog in the metric in curvature time  Eq.~(\ref{1}).

The metric (\ref{1})
has the  following property: It remains unchanged under multiplication
of
 $ a^4$ and $B$ by
 the same positive
constant.
Such a constant factor appears if we
multiply the spatial coordinates by a
constant factor. In synchronized coordinates
this property means that not $a$ itself, but only
the Hubble parameter
\begin{eqnarray}
   H (t): = \frac{\dot a}{a}
\label{10}\end{eqnarray}
has an invariant  meaning. By the same token,  not $B(a)$ itself,
but only ${B(a)}/{a^4}$ has an invariant meaning.
In fact, from Eq.~(\ref{5}) we see that
\begin{eqnarray}
   \frac{B}{a^4}  = H^2.
\label{11}\end{eqnarray}
\section{Cosmological solutions}%%sct. V
Here we  recall some formulas of Ref. \cite{5}, and
present some new results for
 the curvature-saturated  Lagrangian.

\subsection{Solutions for Lagrangian $R^m$}

For the Lagrangian ${\cal L} = R^m$, we obtain  the following exact
solutions for
a closed Friedmann universe:
\be
ds^2 = dt^2 - \frac{t^2}{2m^2 - 2m - 1} d\sigma^2_{(+)},
\ee
where $d\sigma^2_{(+)}$ is the metric of the unit 3-sphere.

Analogously, for the open model
\be\label{@60}
ds^2 = dt^2 - \frac{t^2}{2m - 2m^2 + 1} d\sigma^2_{(+)} .
\ee
Of course, both expressions are valid for positive denominators only.

\medskip

For the spatially flat Friedmann model, it proves useful to employ
the cosmic scale factor $a$ itself
  as a time-like coordinate.
\be\label{@61}
ds^2 = a^2 \left[ Q^2(a) da^2 - dx^2 - dy^2- dz^2
\right].
\ee
This coordinate is
meaningful
 as long as the Hubble parameter is different from zero, so that
we cover only
time intervals where the universe is either expanding or contracting.
Possibly
existing
 maxima or minima of the cosmic scale factor as seen in synchronized time
 can, however, been dealt by a suitable limiting process and
patching.
The curvature scalar reads now
\be\label{@62}
R = \frac{6}{a^3 Q^3} \frac{dQ}{da} ,
\ee
and to reduce the order of the field equation it proves useful to
define
\be\label{@63}
P(a) = \frac{d \ln Q}{da}.
\ee
Then the field equation is fulfilled if
\be\label{@64}
0=m(m-1) \frac{dP}{da} + (m-1)(1-2m)P^2 + m(4-3m) \frac{P}{a}.
\ee
 Therefore, the spatially flat
Friedmann models can be solved in closed form,  but not always
 in synchronized coordinates.

\subsection{Solutions for  Lagrangian ${\cal L}_{\rm cs}$}

In the context of our
curvature-saturated model,
we shall restrict ourselves to the expanding spatially flat
Friedmann model.
The field equation written in synchronized or conformal time---the two most
 often used time coordinates used for this  purpose---have the
disadvantage that the number of terms is quite large, and that even in the
  simplest case ${\cal L} = \frac{1}{2} R^2$ we cannot
give closed-form solutions,
 apart
from the trivial solutions $R \equiv 0$ having the same geometry as
the radiation universe ($a = \sqrt t$ in synchronized time $t$) and
the de
Sitter
universe ($a = e^ t$ in synchronized time $t$). So, we prefer to work in the
less
popular coordinates (\ref{@61}). In principle, the field equation should
be
of fourth order, but we shall reduce it to second order.

To find  the field equation for a spatially flat Friedmann
model  with our Lagrangian, it is
 useful to consider first a general nonlinear
Lagrangian and  specialize to
 ${\cal L}_{\rm cs}$  afterwards. To simplify  (\ref{@62}), we
 define  instead of
 $Q(a)$ the function  $B(a) = Q(a)^{-2} > 0$ as a new dependent function.
Then (\ref{@61}) reads
\be\label{5.4}
ds^2 = a^2 \left[ \frac{da^2}{B(a)} - dx^2 - dy^2- dz^2
\right]
\ee
and  (\ref{@62})
goes over
 to
\be\label{5.5}
R = - \frac{3}{a^3}  \frac{dB}{da} ~  .
\ee
Thus, $B$ itself does not appear explicitly, and only first, and not second
derivatives
are present.
The
geometric
 origin of this property is the same as
in  Schwarzschild
 coordinates---one integration constant is  lost in the definition of
the coordinates,
and this makes curvature depend only on the first derivative of the metric.

\medskip

{}From the 10 vacuum field equations (\ref{4.3}) only the 00-component is
essential;
 it is the constraint equation, therefore it has one order less than the full
field equation, but if the constraint is fulfilled always, then all other
components
 are fulfilled, too.\footnote{This behavior is known already from the
Friedmann equation
in General Relativity: Energy density is proportional to the square of the
Hubble
parameter which contains only a first derivative.} Together with Eq.
(\ref{5.5}) we
should now expect that the fourth order field equation (\ref{4.3}) can be
reduced
to one single second order equation for $B(a)$, where hopefully, $B$ itself no
more
appears.

The
equation $\Theta _{00}=0$ is via (\ref{2.3a}) and (\ref{Theta})
 equivalent to
\be\label{5.6}
0= 3{\cal L}'  \left(2B - a \frac{dB}{da}\right) - a^4 {\cal L} -
18aB \, {\cal L}'' \frac{d}{da}
\left( \frac{1}{a^3}   \frac{dB}{da}  \right) ,
\ee
which is much simpler than the analogous equation in synchronous time, as
observed here
 for the first time.

Before we insert our  Lagrangian ${\cal L}_{\rm cs}$ into  (\ref{5.6}),
let us cross check
 its validity by solving known problems: If ${\cal L}''$ vanishes identically, then ${\cal L}'$
is
a constant, and  we return to Einstein's theory.
The case $B\equiv $ const. gives the
radiation
 universe, while  $B=a^4$ is the exact de Sitter solution. For the
Lagrangian ${\cal L} = \frac{1}{2} R^2$ with
${\cal L}' =R$ and $ {\cal L}''=1$, and Eq.~(\ref{5.6})
reduces to
\be\label{5.7}
0 = a \dot B^2 - 4aB \ddot B + 8 B \dot B,
\ee
 where a dot denotes differentiation with respect to $a$.
Again, $B=a^4$ is the exact de Sitter solution. Defining $\beta = \ln B$ and
 $z = a \dot \beta$, Eq. (\ref{5.7}) goes over in
\be\label{5.8}
4a \dot z = 3z(4-z).
\ee
With $\alpha = \ln a$ we arrive at
\be\label{5.9}
4 \frac{dz}{d\alpha} = 3z(4-z),
\ee
which can be solved in closed form.  Qualitatively it is clear
 that $z=4$, i.e., the de Sitter solution,  represents an attractor.
Solving
 Eq.~(\ref{5.9}) we obtain
in the region $0 < z < 4$:
\be\label{@80}
z = 2 + 2 \tanh \left(\frac{3}{2} \alpha\right),
\ee
showing explicitly that $ z \to 4$ for $\alpha \to \infty$.  The metric can be
calculated from
\be\label{@81}
\dot \beta = \frac{2}{a} \left( 1 + \frac{a^3 - 1}{a^3 + 1} \right),
\ee
using the identity
\be\label{@82}
\tanh \ln x = \frac{x^2-1}{x^2+1}.
\ee

After these preparations we are ready to deal with our Lagrangian
${\cal L}_{\rm cs}$. We insert ${\cal L}$ and ${\cal L}'$
from Eq. (\ref{4.7b}),
 and
\be
 {\cal L}'' = -3l^4 R (1+l^4 R^2)^{-5/2}
\ee
into Eq. (\ref{5.6}) and obtain, after setting $l=1$,
the simple expression
\be\label{5.11}
54 a^9 B \dot B \frac{d}{da}  (a^{-3} \dot B)
 = a^5(a^6 + 9 \dot B ^2)(2B - a \dot B) + \dot B (a^6 + 9 \dot B ^2)^2.
\ee

In these coordinates, the flat Minkowski spacetime does not exist, and the
radiation
 universe $R=0$ is not a solution. This is why $B=$ const. yields
 no solution to Eq. (\ref{5.11}). Also, as was known from the beginning: the
de Sitter
 spacetime $B=a^4$ is not an exact solution here. However, in
the nearby-region where the Lagrangian is well approximated by a quadratic
function in $R$ with a nonvanishing linear term, the behavior of the solutions
is quite similar to that of $R+R^2$-models, where no exact de Sitter solution
exists,
 but a quasi de Sitter solution represents a transient attractor with
sufficient
long duration to solve the known cosmological problems.
 These calculations  have been presented at different places,
most
explicitly
in Ref. \cite{6}. After this phase,  the universe goes to the
weak-field
behavior, where our model behaves as usual.

The main departure of our model from the usual one is in the region of large
curvature scalar, where $ \vert  \dot B \vert $ is large compared to $a^3$.
 To find out the behavior of the solutions in this limit, we compare the
leading terms
in  Eq. (\ref{5.11}) and see that
$ \ddot B$ is proportional to $\dot B^4$, where the
coefficient
of proportionality is positive and slowly varying. Thus, we find
approximately $B(a)\approx a^{2/3}$ for small $a$. This implies the existence
of a big-bang singularity, but with a different behavior: From Eq. (\ref{5.4}) we
obtain
\be\label{@e}
ds^2 = a^2 \left[ \frac{da^2}{a^{2/3}} - dx^2 - dy^2- dz^2
\right],
\ee
which corresponds in synchronized time to the behavior
\be\label{@f}
 ds^2 = dt^2 - t^{6/5}(dx^2 + dy^2+ dz^2),
\ee
this being a good approximation to the exact metric for small $t$,
 differing from the usual big-bang behavior  in almost all other
models. Further details of   our model will be presented elsewhere.

\subsection{The cosmological singularity}%%New subsection VIIC
Here we present the argument with the singularity behaviour mentioned 
 at the end of section II: 
 In our model, differently from
 Einstein's theory, the divergence of the curvature does not necessarily imply the 
 divergence of any part of the energy-momentum-tensor. Let us concentrate on the
 %consideration of the
 trace. The r.h.s. of    Eq. (\ref{@tr}) reads 
$$
\frac{1}{8\pi G}\left\{
\frac{R+ 2l^4 R^3}{\left( 1 + l^4R^2 \right)^{3/2} }-
3  \Box \left[ \frac{1}{\left( 1 + l^4R^2 \right)^{3/2}}
\right]   \right\} 
$$
and this expression must be equal to the trace $T$ of the energy-momentum 
 tensor. In Einstein's theory, $R \to \pm \infty $ necessarily implies 
 $T \to \pm \infty$, whereas here, $T$ may remain finite even if $R \to \infty$.

Detailed numerical calculations would  support this qualitative 
picture, however, we  postpone such calculations until we have a more
 strictly physically motivated form of the Lagrangian.

\section{Discussion}%%VI

 We have argued that the   gravitational action ${\cal A}$
has a decreasing dependence on $R$
for increasing
$\vert R \vert $.
Such a behavior is expected from the spacetime stiffness
caused by the vacuum fluctuations
of all quantum fields in the universe.

Our model does not have the tachyonic disease
of $R+R^2$ models
studies by
Stelle \cite{13} and others
 \cite{14}.

Since our model has an action which interpolates between
Einstein's action and a pure cosmological term,
it promises to have interesting observable consequences
which may explain some of the experimental cosmological data.

%%ZZ\comment{
The heat-kernel expansion of the effective action in a curved background
is closely related to the  Seeley-Gilkey coefficients \cite{15}, and for higher
 loop expansion also higher powers of curvature appear: To get the  $n$-loop
 approximation
one has to add terms until $\sim R^{n+1}$, a behavior which also
 happens in the string effective action \cite{16}. So, if one cuts this
 procedure at a certain value of $n$, one gets always as leading term for high
curvature
 values a term like  $\sim R^{n+1}$. However, the $n$-loop approximations need
not
 converge to the correct result if one simply takes $n \to \infty$ in the
$n$-loop-result. In fact, what we have used in the present paper is such an
example:
\be\label{7.1}
{\cal L}_{\rm cs} = \frac{R}{16 \pi G \sqrt{1+l^4 R^2}}=  \frac{R}{16 \pi G} +
\sum_{k=1}^{\infty} ~b_k ~R^{2k+1}
\ee
 with some real constants $b_k$, where
\be\label{7.2}
b_1 = - \frac{l^4}{32 \pi G}
\ee
but the Taylor expansion on the right hand side diverges for $R > l^{-2}$.
So, the Taylor expansion is useful for small $R$-values only, and for large
values $R$ we need a correct analytical continuation.
%%ZZ}

Prigogine et al. 
have proposed
in Eq. (18) of
Ref.~\cite{17}
 a model where the
 effective gravitational constant   depends on the Hubble parameter
 of a Friedmann model. Though this ansatz depends on the special
3+1-decomposition of spacetime, it shares some similarities with the
 model discussed here. More recent developments how to find a
well-founded gravitational action from considering quantum effects
can been found in \cite{18} and \cite{19}.

Quite recently, see for instance \cite{22},
accelerated expansion models of the
universe have been  discussed and compared with new observations. We  postpone
 the comparison of our model with these observations to  later work.

\section*{Acknowledgment}
H.-J. S. gratefully acknowledges financial support from DFG
 and from the HSP III-program.
We thank V. Gurovich and  the colleagues of the Free University Berlin,
where this work has been done, especially M. Bachmann
 and A. Pelster, for valuable comments.

\appendix
\section{Newtonian limit in a nonflat background}%%App. A
The Newtonian limit of a theory of gravity
is defined as follows: It is
the weak-field slow-motion limit for fields whose
energy-momentum tensor is dominated
by its zero-zero component in comoving time.
Usually, the limit is formed
in a flat background, and sometimes,
this is assumed to be a necessary assumption. This is, however, not true, and
we
show here briefly how to calculate the Newtonian limit in a nonflat
background,
Moreover, our approach is different from what is usually called
Newtonian cosmology.
%thus differing from the usual Newtonian
%cosmos).
 To have a concrete example, we take the
background as a de Sitter spacetime.

The slow-motion assumption allows us
 to work with
static spacetime and the matter,
assuming
the energy-momentum tensor
to be
\begin{equation}
T_{ij} =  \rho ~ \delta ^0_i  \delta ^0 _j,
\label{@X7}\end{equation}
 where $ \rho $ is the energy density, and time is assumed to be synchronized.
The de Sitter spacetime in its static form can be given as
\begin{eqnarray}
 ds^2 = - ( 1 - kr^2) dt^2 + \frac{dr^2}{1 - kr^2}
 + r^2 d  \Omega ^2                        ,
\label{A.1}\end{eqnarray}
 where $ x^0 = t,~ x^1 = r,~ x^2 = \chi,~ x ^3 = \theta$
and
$d  \Omega ^2 = d\chi^2 + \sin^2 \chi d \theta^2$
is the metric of the 2--sphere.
In this Appendix, we have changed the
signature of the metric
from $(+---)$,
which is usual in cosmology,  to $(-+++)$,
which leads to the standard definition of the
Laplacian.

The parameter $k$ characterizes  the following physical situations: For $k =
0$,
 we
have the usual flat background. By setting
$k = 0$ we can therefore compare the results
with the well-known ones.
The case $k > 0$ corresponds to a positive cosmological constant $ \Lambda $.
In the calculations, we must
observe that the  time coordinate $t$
 fails to be a synchronized for $k \ne 0$, but it is obvious from the
context how to obtain the synchronized time from it.

In the coordinates (\ref{A.1}), there is a horizon at
$ r = r_0 \equiv \frac{1}{ \sqrt{k} }$. So, our approach
makes sense in the interval $ 0 < r < r_0$. However,
 $r_0$ shall be quite large in comparison with the system
under consideration, so that we do not meet a problem here.

Now, the following ansatz seems appropriate:
\begin{eqnarray}
 ds^2 = - ( 1 - kr^2) ( 1 - 2 \varphi) d t^2 +
 \left( \frac{dr^2}{1 - kr^2} + r^2 d  \Omega ^2 \right)
 ( 1 + 2 \psi),
\label{A.2}\end{eqnarray}
 where $\varphi$ and $\psi$ depend on the spatial coordinates only.
The weak-field assumption allows us to make linearization with
respect to $\varphi$ and $\psi$.
An extended matter configuration can be obtained
by superposition of point particles, so we only
need to solve the problem for a $ \delta $-source
at $r = 0$. This one is spherically symmetric, so we may
assume $ \varphi = \varphi (r)$ and
$\psi = \psi (r)$ in Eq.~(\ref{A.2}).
For the metric components we get:
\be
g_{00} = - ( 1 - kr^2) (1 - 2 \varphi),~~~~
g_{11} = \frac{1 +2\psi}{1 - kr^2},~~~~
g_{22} = r ^2 ( 1 + 2 \psi),~~~~
g_{33} = g_{22} \cdot \sin^2  \chi         .
\ee
The inverted components are up to linear order in
$\varphi$ and $ \psi$:
\be
g^{00} = - \frac{1+2\varphi}{1 - kr^2},~~~~
 g^{11} = ( 1 - kr^2)(1 - 2 \psi),~~~~
g^{22} = \frac{1 - 2\psi}{r^2},~~~~
g^{33} = g^{22} \sin^{-2} \chi,
\ee
which gives the Christoffel symbols
\begin{eqnarray}
 \Gamma _{01}^0 &=& - \varphi' - \frac{kr}{1 - kr^2},\\
 \Gamma _{00}^1& =&  ( 1 - kr^2) \left[ - kr + 2 kr
  (\varphi + \psi) - \varphi' (1 - kr^2)\right],\\
  \Gamma ^1_{11} &=& \psi' + \frac{kr}{1 - kr^2}
,\\
 \Gamma _{12}^{2} &=&  \Gamma ^3_{13} = \psi' + \frac{1}{r}
,\\
\Gamma _{22}^{1} &=& - r ( 1 - kr^2) - \psi' r^2 ( 1 - kr^2)
,\\
  \Gamma _{33}^1 &=& \sin^2 \chi   ~ \Gamma _{22}^1
,\\
 \Gamma _{32}^3 &=& \cot \chi
,\\
 \Gamma _{33} ^2 &=&  - \sin \chi \cos \chi,
\label{@MA}\end{eqnarray}
and the Ricci tensor reads
\begin{eqnarray}
R_{00} &=& - 3 k ( 1 - kr^2) - \varphi '' ( 1 - kr^2)^2 -
 \frac{2 \varphi'}{r} ( 1 - kr^2 )
+ 6 k ( \varphi + \psi) ( 1 - k r^2) + kr ( 1 - kr^2)
 (5 \varphi' - \psi'),\\
R_{11} &=& - 2 \psi '' + \varphi'' - \frac{2}{r} \psi' + \frac{3k}{1-kr^2}
+  \frac{k r}{1-kr^2} (\psi ' - 3 \varphi'),\\
R_{22}& =& 3k r^2 - \psi'' r^2 ( 1 - k r^2 )   - \psi'  (2r - 4k r^3 )
+  (\varphi' -  \psi')( r - k r^3),\\
R_{33} &=& R_{22} \cdot  \sin^2 \chi   .
\label{@RA}\end{eqnarray}
Before we discuss these equations, we consider two obvious limits:
\bigskip

For $k = 0$, we see that $R_{00} = - \varphi''
 - {2 \varphi  '  }/{r} = -  \Delta  \varphi$, leading to the usual
 Newtonian limit $\Delta \varphi = - 4 \pi G \rho$.

\bigskip

For $\varphi = \psi = 0$ we get for the Ricci tensor:
\be
R^0_0 = R^1_1 = R^2_2 = R_3^3 = 3k,
\ee
and thus the de Sitter spacetime with $R=12k$ for $k >0$.

\bigskip

Returning to the general case we have
\be
\frac{R}{2} = 6k - 12k \psi + (\varphi  '' - 2 \psi   '' ) (1-kr^2) +
\frac{2}{r} \varphi  '  - 5kr \varphi  '  - \frac{4}{r} \psi  ' + 7kr \psi  '
\ee
and then
\be
R^0_0 - \frac{R}{2} = - 3k + 6k\psi + 2 \psi '' (1-kr^2) -
 6kr \psi  ' + \frac{4}{r} \psi  '.
\ee
The other components have a similar structure and can be
calculated easily from the above equations.
The first term of the r.h.s., $-3k$, will be compensated by the $\Lambda$-term.
The usual gauging to $\psi \to 0$ and $\varphi \to 0$ as $r \to \infty$ is no
more
possible because for $r > r_0$ our approximation is no more valid. As an
 alternative gauge we add such constant values to $\psi$ and $\varphi$ that
 they are approximately zero in the
region under consideration.  So we may disregard the
term
 $6k\psi$. All remaining terms with $k$ can be
obtained from those without $k$ by multiplying with factors of the type
 $1 + \epsilon$ where $\epsilon \approx k r^2$, $k=1/r_0^2$, with
 $r_0$ being of the order of magnitude of the world radius. In a first
approximation, this gives only a small correction to the gravitational
constant. In a second approximation, there are deviations from the $1/r$-behavior.

An
 analogous discussion for the Lagrangian
$R + l^2 R^2$ tells us that
in a range where $l \ll r \ll  r_0$, the
potential behaves like $(1 - c_1 e^{-r/l} )/r$, as in flat space.

\section{The absence of ghosts and tachyons} %%New App. B
Here we show in more details what has been stated after Eq. 
(\ref{@11}).  In the conformally transformed picture with a scalar field, the 
absence of tachyons (i.e., particles with wrong sign in front of the 
potential term)  becomes clear from the form of  the  potential.  
 For checking ghosts (i.e., particles with wrong sign in front of the kinetic term)
 we have to go a little more into the details: 
%\be\label{Z1}
%y=x
%\ee
%Like in eq. (\ref{Z1}).
 In Stelle \cite{23} the particle content of fourth order gravity with terms
 up to quadratic order has been determined, and the existence/absence of ghosts
 and tachyons has   been given in dependence on the free constants of the theory.  
 In the first of Refs. \cite{4}, the analogous calculation as in \cite{23} has been 
done for a  term $R^3$ added to the Einstein--Hilbert-Lagragian. Let us give here the
 argument for general $n \ge 3$: If $R^{n}$ is in ${\cal  L}$, then the term
 $R^{n-1}$ and its derivatives are in the corresponding expression after 
variational derivative with respect to the metric. In the result, all  terms
 represent products of at least $n-1$ 
 small quantities; because of $n \ge 3$ these  are always at least two factors; 
 thus, they all vanish in the
 linearization about the Minkowski space--time.

\bigskip

Now, one might be tempted to require the analogous  linearization
properties for a Friedmann--Robertson--Walker background. However, 
linearization around other than flat space--times is not at all a trivial task, see \cite{25},
 even for Einstein's  theory: For the closed Friedmann model, Einstein's theory 
 is 
linearization unstable, for spatially flat models it is stable, and for the open
 Friedmann model the result  is -- contrary to other claims in the  older
 literature -- not yet known. We face the further problem
 that linearization around
 the de Sitter space-time is  complicated to determine, because the same geometry
 can be locally represented as a spatially flat as well as a closed Friedmann model. 
 So, we leave  the question of linearization stability with non-flat background of our 
model unanswered. 

\bigskip

Another type of reasoning was given quite recently: 
In \cite{24} the possibility  has been discussed  that  the contributions 
 to the Lagrangian coming  of gravitons on the one hand 
and of gravitinos on the other 
 may cancel each other  to avoid  the ghost problem.

\end{document}